\documentclass{article}

\usepackage{PRIMEarxiv}

\usepackage[utf8]{inputenc} 
\usepackage[T1]{fontenc}    
\usepackage{hyperref}       
\usepackage{url}            
\usepackage{booktabs}       %
\usepackage{siunitx} 
\usepackage{amsfonts}       
\usepackage{nicefrac}       
\usepackage{microtype}      
\usepackage{lipsum}
\usepackage{fancyhdr}       
\usepackage{graphicx}       
\graphicspath{{media/}}     
\usepackage{graphicx}
\usepackage{subcaption}

\usepackage[toc,page]{appendix}
\usepackage{lineno}
\usepackage{algorithm}

\usepackage{algpseudocode}
\usepackage{amsmath}
\usepackage{makecell}
\usepackage{multirow} 
\usepackage[colorinlistoftodos,prependcaption,textsize=tiny]{todonotes}

\title{Precise Antigen-Antibody Structure Predictions Enhance Antibody Development \\with HelixFold-Multimer
}

\author{
 Jie Gao\thanks{Equal contributions.}, Jing Hu\footnotemark[1], Lihang Liu, Yang Xue, Kunrui Zhu, Xiaonan Zhang,  Xiaomin Fang\thanks{Corresponding author. Email: fangxiaomin01@baidu.com } \\
 PaddleHelix team, Baidu Inc. \\
 }

\begin{document}
\maketitle

\begin{abstract}
The accurate prediction of antigen-antibody structures is essential for advancing immunology and therapeutic development, as it helps elucidate molecular interactions that underlie immune responses. Despite recent progress with deep learning models like AlphaFold and RoseTTAFold, accurately modeling antigen-antibody complexes remains a challenge due to their unique evolutionary characteristics. HelixFold-Multimer, a specialized model developed for this purpose, builds on AlphaFold-Multimer’s framework and demonstrates improved precision for antigen-antibody structures. HelixFold-Multimer not only surpasses other models in accuracy but also provides essential insights into antibody development, enabling more precise identification of binding sites, improved interaction prediction, and enhanced design of therapeutic antibodies. These advances underscore HelixFold-Multimer's potential in supporting antibody research and therapeutic innovation.

\end{abstract}


\section{Introduction}
The precise prediction of antigen-antibody structures is critical for immunology and therapeutic development. Accurate modeling of these structures underpins understanding of molecular interactions that drive immune recognition and response and is vital for the development, optimization, and design of antibodies for effective antibody-based vaccines and therapies. 

Recent advancements in computational biology have introduced groundbreaking methods that markedly enhance the accuracy of protein structure prediction. State-of-the-art deep learning-based models \cite{jumper2021highly, evans2021protein,abramson2024accurate,baek2021accurate,krishna2024generalized,lin2023evolutionary,fang2023method}, represented by the AlphaFold \cite{jumper2021highly, evans2021protein,abramson2024accurate} and RoseTTAFold \cite{baek2021accurate,krishna2024generalized} series, have made substantial progress in the end-to-end prediction of single protein chains, protein complexes, and biomolecular interactions. These models utilize evolutionary information derived from multiple sequence alignments (MSA) to predict pairwise residue-residue evolutionary and spatial relationships, which are crucial for accurate structural modeling. 
Despite these advances, the precision of antigen-antibody structure prediction remains suboptimal \cite{yin2022benchmarking,yin2024evaluation}. This is likely due to the distinct evolutionary patterns of antigens and antibodies compared to conventional protein complexes. Methods like AFSample \cite{wallner2023afsample} and AlphaFold3 \cite{abramson2024accurate}) attempt to employ extensive structure sampling to improve prediction accuracy, but the computational demands of conducting hundreds or thousands of sampling times are often prohibitive, especially for high-throughput applications.

Specialized methods \cite{ruffolo2023fast,ruffolo2022antibody,leem2016abodybuilder,sircar2009rosettaantibody,abanades2022ablooper}, such as IgFold \cite{ruffolo2023fast}, DeepAb \cite{ruffolo2022antibody}, and AbodyBuilder \cite{leem2016abodybuilder}, have been developed to improve antibody structure prediction.
These methods have demonstrated improved accuracy, especially in the complementarity-determining regions (CDRs), which are crucial for antibody function. These advances hold significant promise for applications in antibody development and therapeutic development, particularly in predicting the properties and functions of mutated antibodies. Nevertheless, these approaches primarily focus on predicting antibody structures alone and do not extend to accurately predicting complex interactions between antigens and antibodies. Consequently, the accurate prediction of antigen-antibody structures remains an ongoing challenge.

The rapid advancements in predicting protein-protein interactions, optimizing protein properties, and designing novel proteins with specific attributes are showing significant promise across various research domains. Protein structure models are essential in these applications, as they rely on the detailed structures or scoring metrics generated by protein structure prediction models to accurately forecast protein-protein interactions and guide the design of new proteins.

Mainstream deep learning-based methods for predicting protein-protein interactions (PPIs) typically fall into three categories: sequence-based models, which utilize protein sequence data (i.e., residue sequences) \cite{du2017deepppi,xie2020prediction,patel2017deepinteract}; structure-based models, which rely on three-dimensional protein structures \cite{singh2010struct2net,northey2018intpred}; and hybrid models that integrate both sequence and structural information \cite{xue2022multimodal}. These methods are designed to identify interaction sites or determine the likelihood of protein binding.
Recent developments have shown that protein complex structure prediction models are particularly effective for PPI prediction. By analyzing the structures of predicted protein complexes, these models can provide direct insights into interaction sites and binding patterns. Furthermore, the use of AlphaFold’s varying confidence scores \cite{bryant2022improved,bennett2023improving,kim2024enhanced} has proven effective in distinguishing between interacting and non-interacting proteins. Additionally, newly developed screening pipelines \cite{humphreys2021computed,yu2023alphapulldown} enable the analysis of large-scale protein datasets, aiding in the discovery of novel PPIs. These developments highlight the substantial potential of structural prediction models to enhance our understanding of protein-protein interactions and their biological implications.

Many cutting-edge studies employ deep learning models to optimize proteins through mutations or to design entirely new proteins with specific structural or functional requirements. These models, known as protein language models, fall into two main categories: sequence-based models (e.g., ESM \cite{rives2021biological, lin2022language}, ProGen \cite{madani2023large, nijkamp2023progen2}, and ProtGPT2 \cite{ferruz2022protgpt2}) and structure-based models, also known as inverse folding models \cite{hsu2022learning,dauparas2022robust,shanker2024unsupervised,cheng2024zero} (e.g., ESM-IF \cite{hsu2022learning} and ProteinMPNN \cite{dauparas2022robust}). Research utilizing ProteinGym \cite{notin2023proteingym} has highlighted the significant potential of these models to enhance critical protein attributes, including activity, binding affinity, expression levels, organismal fitness, and stability. In particular, structure-based models generally provide higher accuracy compared to sequence-based models by using structural data from reference proteins. This data can include crystallized structures or, more frequently, structures predicted by high-precision structural prediction models, as many reference proteins lack crystallized structures.

Despite considerable progress in applying protein structure prediction models to protein-protein interactions (PPI), protein optimization, and design, research has primarily focused on general proteins, with relatively few studies addressing the complexities of antibody-specific applications. This shortfall is largely attributed to the insufficient accuracy of current structure prediction models in accurately capturing the structures and scoring metrics of antigen-antibody interfaces. As a result, the integration of deep learning techniques into antibody research remains limited. To bridge this gap, it is crucial to develop and rigorously evaluate high-precision structure prediction models tailored specifically for antigen-antibody interactions, with the potential to significantly enhance antibody development.

To this end, we developed a specialized structure prediction model for antigen-antibody interactions, named HelixFold-Multimer. This model builds upon the training framework of AlphaFold-Multimer \cite{evans2021protein} but has been specifically adapted for antigen-antibody systems, fine-tuning with the antibody data. Our results show that HelixFold-Multimer significantly outperforms competitive baseline methods, such as AlphaFold-Multimer \cite{evans2021protein} and AlphaFold3 \cite{abramson2024accurate}, in predicting antigen-antibody structures with superior accuracy. Furthermore, we have demonstrated that high-precision structure prediction is crucial for accurately identifying binding sites, predicting interactions, and optimizing and designing antibodies. HelixFold-Multimer not only enhances prediction accuracy and expands application scope when used with other protein development tools, but also provides valuable insights through its confidence scores and mask language model evaluations, which can be directly utilized to improve antibody screening and design. These advances highlight the significant potential of the model in advancing antibody development and contribute to the broader field of structural biology.

\section{Antigen-Antibody Structure Prediction}

\begin{figure}
    \centering
    \includegraphics[width=1.0\linewidth]{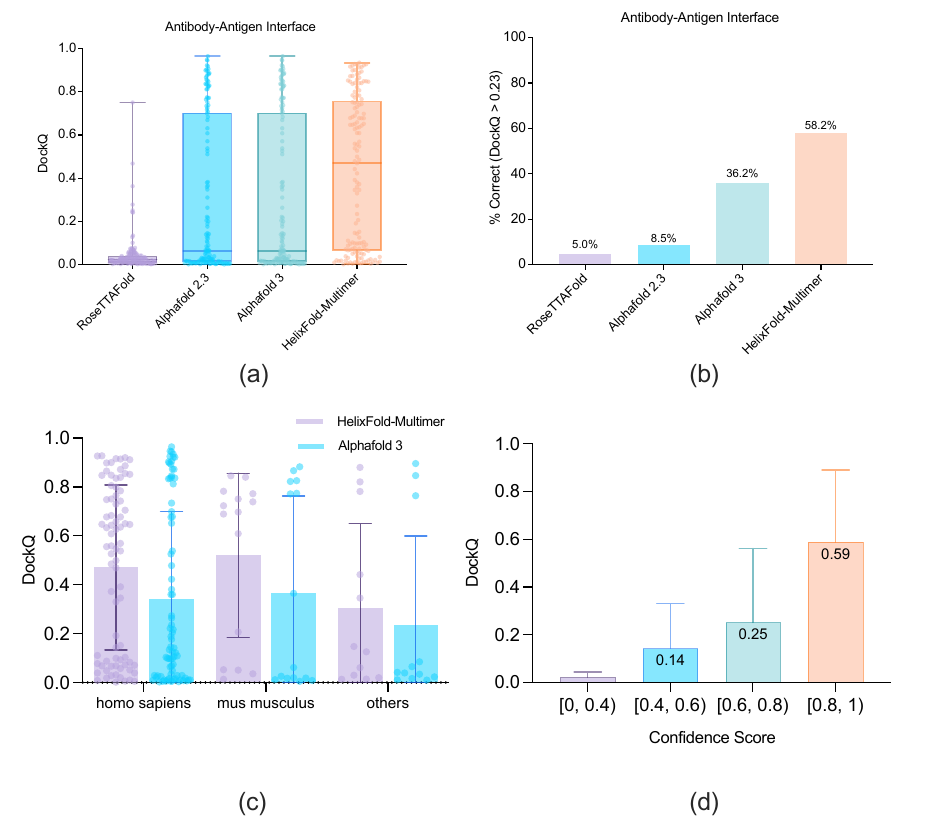} 
    \caption{Performance comparison of HelixFold-Multimer and the leading structural prediction models for antibody-antigen interfaces. (a) Median DockQ scores. (b) Percentage of predictions exceeding a DockQ threshold of 0.23. (c) Performance of HelixFold-Multimer and AlphaFold 3 across different species. (d) Relations between the DockQ scores of HelixFold-Multimer prediction and confidence levels outputted by HelixFold-Multimer.
    }
    \label{fig:fig_structure_main}
\end{figure}

HelixFold-Multimer builds upon the AlphaFold-Multimer \cite{evans2021protein} architecture with small adjustments in the structure module. While the modeling of residue-level interactions is retained, it also incorporates the modeling of chain-level interactions to enhance overall accuracy. HelixFold-Multimer first follows the data setting of AlphaFold-Multimer, training a general version model with protein data from the Protein Data Bank (PDB) and self-distillation data, achieving competitive results compared to AlphaFold-Multimer. The model is then fine-tuned with antigen-antibody structures collected from the PDB to obtain a specialized model specifically tailored for predicting antigen-antibody interactions. 
For antibody sequences, we additionally searched the Antiref database to enrich the multiple sequence alignment (MSA) data.

\subsection{Overall Accuracy}
To evaluate the precision of HelixFold-Multimer on antigen-antibody structure prediction, we collected 141 recently released antigen-antibody interfaces from the Protein Data Bank (PDB) (refer to Section \ref{dataset_antibody} for details on data preprocessing). We benchmarked HelixFold-Multimer against leading models, including RoseTTAFold \cite{baek2021accurate}, AlphaFold-Multimer \cite{evans2021protein}, and AlphaFold3 \cite{abramson2024accurate}.
As illustrated in Figs.~\ref{fig:fig_structure_main}(a) and (b), HelixFold-Multimer outperformed all competitors, achieving a median DockQ score of 0.469 and a success rate of 58.2\%, significantly exceeding AlphaFold3's median DockQ score of 0.065 and a success rate of 36.2\%.
We also presented the performance of the general version of HelixFold-Multimer (Supplementary Figure \ref{fig:fig_structure_general}), which lacks specific optimizations for antigen-antibody scenarios. The results demonstrated a notable decline in accuracy, underscoring the effectiveness of the targeted enhancements in improving the model's ability to predict antigen-antibody complex structures.

\subsection{Accuracy across Antibody Species}
Antibodies across species differ in their isotypes, constant region sequences, and glycosylation patterns. These structural differences significantly affect their stability, specificity, and functionality. To investigate the effectiveness of HelixFold-Multimer in predicting antibody structures across different species, we categorized the species into three groups based on the origin of the antibody heavy chains. Homo sapiens, Mus musculus, and other species, including Macaca mulatta, Rattus norvegicus, synthetic constructs, Gallus gallus, and Oryctolagus cuniculus. As illustrated in Figs.~\ref{fig:fig_structure_main}(c), both HelixFold-Multimer and AlphaFold3 demonstrate higher accuracy for Homo sapiens and Mus musculus compared to other species. This result is consistent with expectations, as the majority of antibody data are derived from studies on Homo sapiens and Mus musculus. HelixFold-Multimer consistently outperforms AlphaFold3 across all species groups, with particularly significant improvements observed for the extensively studied Homo sapiens and Mus musculus.  These enhancements could benefit the development of antibody-based therapeutics.

\subsection{Precision of the Confidence Scores}

The confidence measures of HelixFold-Multimer are closely aligned with the accuracy of the model. We assessed the correlation between DockQ scores and three confidence metrics: confidence scores, ipTM scores, and pLDDT scores. All metrics demonstrated a strong positive correlation with DockQ (Pearson correlation: 0.664 for confidence scores, 0.658 for ipTM, and 0.344 for pLDDT), with confidence scores and ipTM showing particularly strong associations. Figure~\ref{fig:fig_structure_main}(d) displays the distribution of DockQ scores across different confidence score ranges. These confidence measures offer valuable insights for the effective application of HelixFold-Multimer in antibody development.

\section{Epitope-Specific Antigen-Antibody Structure Prediction}

\begin{figure}
    \centering
    \includegraphics[width=1.0\linewidth]{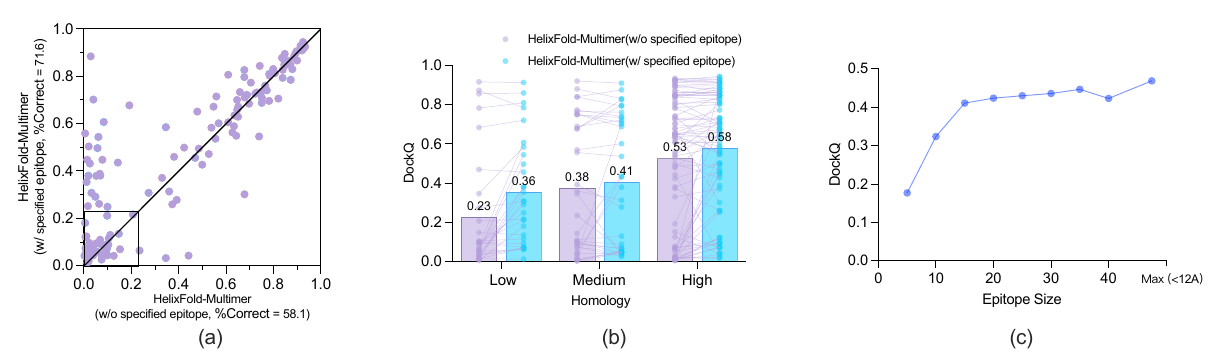} 
    \caption{ Impact of Epitope Information on Antibody-Antigen Docking Accuracy. (a) Scatter plots comparing DockQ scores for HelixFold-Multimer predictions with and without epitope inclusion. (b) The DockQ bar plots of antigens with different homology. (c) The influence of epitope size on the prediction accuracy of antigen-antibody interactions.}
    \label{fig:fig_pocket}
\end{figure}

Accurate antigen-antibody structure prediction depends significantly on the inclusion of antigen epitope information. When epitope residues are specified, HelixFold-Multimer enhances the interaction modeling between antibody and antigen residues by refining attention mechanisms within the EvoFormer and Structure Module.
During training, antigen residues within a threshold distance from the antibody were designated as epitopes. In practical scenarios, experimental techniques such as deep mutational scanning are often utilized to obtain epitope information, which can enhance the accuracy of structural predictions by the model, thereby aiding in drug development.

\subsection{Accuracy Improvement with Specified Epitopes}
Specifying antigen epitope locations allows HelixFold-Multimer to produce more accurate structural predictions. When epitope locations were provided, antibodies that initially docked incorrectly to the antigen binding site were able to dock correctly, as demonstrated by a significant increase in DockQ and success rate (Figure~\ref{fig:fig_pocket}(a)).
Figure~\ref{fig:fig_pocket}(b) illustrates the impact of specifying epitope location on HelixFold-Multimer's performance across different antigen types. The test set was categorized into three groups based on antigen sequence identity with the training set: low homology (sequence identity below 40\%), medium homology (sequence identity between 40\% and 95\%), and high homology (sequence identity above 95\%). The results demonstrate that specifying epitope locations improved HelixFold-Multimer's performance across all groups. The most substantial improvement in prediction accuracy was seen for low homology antigens. The model encounters greater difficulty with these antigens; however, specifying epitope locations mitigates these challenges and enhances predictive accuracy.


\subsection{Effect of Epitope Completeness}
We further examined the impact of epitope completeness on the model's inference performance. Epitopes were defined as a subset of antigen residues in closest proximity to the antibody, identified based on spatial distance, under the assumption that these residues are most likely to participate in antigen-antibody interactions. The analysis (Figure~\ref{fig:fig_pocket}(c)) reveals that as the number of selected residues increases, the model's DockQ score consistently improves. This trend suggests that providing a more complete set of epitope residues enhances the model’s ability to accurately predict the antigen-antibody complex structure.
However, when the number of epitope residues is very small (e.g., 5 or 10), there is a notable decrease in the DockQ score, sometimes even falling below the accuracy achieved without any specified epitope information. This decrease is likely due to the model's tendency to overly focus on a limited set of residues, which restricts its ability to learn from the broader structural context of the antigen-antibody interaction.
Therefore, selecting an optimal number of epitope residues is crucial. It ensures a balance between providing sufficient interaction sites to improve model accuracy and avoiding overemphasis on specific regions that could reduce the model's overall predictive power.

\section{Antigen-Antibody Interaction Prediction}

\begin{figure}
    \centering
    \includegraphics[width=1.0\linewidth]{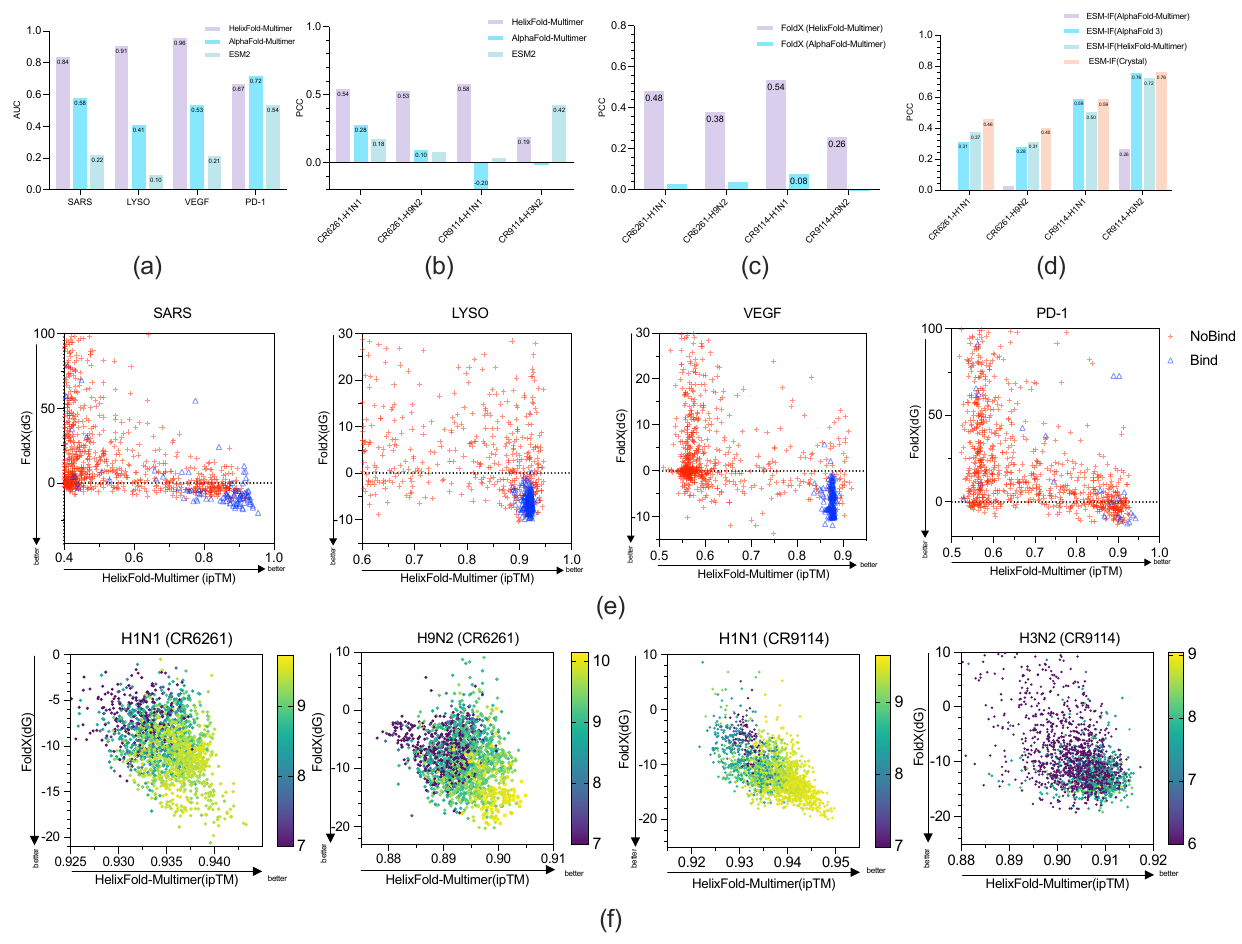} 
    \caption{Comprehensive evaluation of HelixFold-Multimer's performance in antigen-antibody interaction prediction.
(a) Binder recognition capacity for HelixFold-Multimer confidence metric across four antigens.
(b) Affinity prediction capacity for HelixFold-Multimer confidence metric across various influenza strains.
(c) Affinity prediction capacity for energy-based method FoldX using the HelixFold-Multimer or AlphaFold-Multimer predicted conformations as input.
(d) Correlation between ESM-IF scores using input structures derived from AlphaFold3, HelixFold-Multimer, and crystallographic structures, and the experimentally measured affinities.
(e) Integrated
 enhancement of HelixFold-Multimer and energy-based method for binder recognition task.
(f) Integrated
 enhancement of HelixFold-Multimer and energy-based method for binding affinity prediction task. The color bar indicates the experimental binding affinity of various antibody-target pairs.
    }
    \label{fig:fig_inter_main}
\end{figure}

Accurate prediction of antigen-antibody interactions is fundamental to antibody development and relies on the structural understanding of antigen-antibody complexes. We assess the contributions of HelixFold-Multimer's confidence metrics and predicted structures to the accuracy of antigen-antibody interaction predictions.
We first assess the ability of HelixFold-Multimer's predicted structures and confidence scores to inform accurate antigen-antibody interaction predictions. Next, we evaluate the potential performance improvements achieved by combining scores predicted by HelixFold-Multimer with those outputted by energy-based methods.

\subsection{Confidence Metrics of HelixFold-Mutlimer for Interaction Prediction}
The confidence metrics of the AlphaFold-style models were originally designed to assess the reliability of the predicted structures. However, recent studies have shown that these metrics can also be used to estimate protein-protein interactions.
Building on this idea, we explore whether the confidence metric, ipTM, generated by HelixFold-Multimer, can effectively estimate antigen-antibody interactions. 

The confidence metric of HelixFold-Multimer is first evaluated for binder recognition. Binders for SARS-CoV-2 (RBD domain), lysozyme, VEGF (vascular endothelial growth factor), and PD-1 (programmed cell death protein 1) were obtained from the Observed Antibody Space (OAS) for SARS-CoV-2 [\cite{olsen2022observed}], the anti-Lysozyme dataset for lysozyme [\cite{chungyoun2023flab}], the VEGF dataset for VEGF [\cite{chungyoun2023flab}], and Thera-SAbDab for PD-1 [\cite{raybould2020thera}], respectively. In addition, 1000 samples were randomly selected from the B cell receptor (BCR) sequencing data of a healthy donor to serve as non-binders.
Figure~\ref{fig:fig_inter_main}(a) presents a comparison of the area under the curve (AUC) scores, which reflect the quality of ranking, for the confidence metric of HelixFold-Multimer, the confidence metric of AlphaFold 2.3 \cite{yin2022benchmarking}, and the sampling score generated by ESM2 \cite{lin2023evolutionary} across four antigen targets. HelixFold-Multimer achieves AUC scores greater than 0.8 for SARS, LYSO, and VEGF, significantly outperforming the baseline methods, while its AUC for PD-1 is on par with the baseline methods. These results demonstrate that the confidence metric of HelixFold-Multimer exhibits strong ranking performance in the binder recognition task.
Compared to overall ranking ability, determining whether the top-ranked candidates are true binders is more critical for antibody virtual screening. Therefore, the enrichment factor (EF) scores, a widely used metric in virtual screening, are also reported in Suppl. Table~\ref{tab:bind_compare_method}. HelixFold-Multimer demonstrates a high EF$^{5\%}$ for most antigens, indicating its strong performance in prioritizing true binders.
We next assess the ability of the HelixFold-Multimer confidence metric to predict antibody affinity for various influenza strains (H1N1, H3N2, and H9N2) when paired with different antibody variants, such as CR6261 and CR9114 \cite{phillips2021binding}. 
To provide antigens for different influenza strains as model inputs, we first searched the RCSB PDB structure database \cite{zardecki2022pdb} for antigen structures corresponding to these strains. For H1N1, we identified the structure with PDB ID 3GBN, and for H3N2, we used the structure with PDB ID 4FQY. However, for the H9N2 subtype, no corresponding structural data was available. As a result, we turned to sequence databases \cite{uniprot2023uniprot} and used the raw H9N2 sequence as the antigen input for the model.
As shown in Figure~\ref{fig:fig_inter_main}(b), the confidence metrics provided by HelixFold-Multimer exhibit Pearson correlation coefficients greater than 0.5 for three of the four tested antigens, indicating a notable improvement over alternative models. Additionally, Suppl. Figure~\ref{fig:fig_dg_affinity} visualizes the relationship between the predicted confidence scores and the experimentally measured binding affinities for each antigen-antibody pair.


\subsection{High-Precision Structures of HelixFold-Multimer for Superior Interaction Prediction}

Structure-based protein-protein interaction prediction methods, including energy-based methods and deep learning-based methods, often depend on experimentally determined co-crystal structures as input. However, the significant cost and technical challenges associated with obtaining these structures limit the scalability and applicability of such approaches. Using predicted antigen-antibody structures as alternative inputs provides a practical and cost-effective solution to address these limitations.

Energy-based methods, e.g., FoldX \cite{schymkowitz2005foldx}, Rosetta \cite{barlow2018flex}, MMGBSA, and MMPBSA\cite{genheden2015mm} calculate the energy of the interaction between protein pairs by evaluating various energy terms, such as van der Waals forces, electrostatic interactions, and solvation effects to estimate the likelihood of binding. 
To assess the relationship between energy-based methods and conformation quality, we analyzed the correlation between experimentally measured binding affinities and interaction energies calculated by FoldX, using conformations predicted by different structure prediction methods (Figure~\ref{fig:fig_inter_main}(c)). 
When using conformations predicted by AlphaFold-Multimer as inputs for FoldX, the correlation with affinity data was minimal. However, significant correlations were observed when HelixFold-Multimer-predicted conformations were used, underscoring the critical role of accurate conformational predictions in enhancing the reliability of energy-based binding affinity evaluations.

Structure-informed models, such as ESM-IF \cite{hsu2022learning} and ProteinMPNN \cite{dauparas2022robust}, have been shown in previous studies \cite{shanker2024unsupervised}\cite{notin2023proteingym}\cite{hoie2024antifold} to effectively predict protein-protein interactions. These models use the structures of the antigen and reference antibodies to estimate the binding affinities of mutated antibodies. We assess the influence of the structural accuracy of the antigen-(reference) antibody complex on the performance of these models. Specifically, we examine the correlation between experimentally measured affinities and the affinity scores predicted by ESM-IF, using input structures generated by different protein folding methods, as shown in Figure~\ref{fig:fig_inter_main}(d). The DockQ scores of the predicted structures from these methods are provided in Supp. Table~\ref{tab:dockq_inverse_folding_dg}. 
We followed the approach outlined in \cite{shanker2024unsupervised}, where evaluation was performed by providing the entire antibody variable region and antigen complex (Ab-Ag). Antibody sequences were scored by the structure-informed language model with antigen information, using input complexes of CR9114 with H5 HA [PDB ID 4FQI \cite{dreyfus2012highly}] and CR6261 with H1 HA [PDB ID 3GBN \cite{ekiert2009antibody}]. 
As expected, ESM-IF shows the best performance when crystallographic structures were provided as input. The model's performance is slightly lower when using high-precision structures predicted by AlphaFold3 and HelixFold-Multimer. In contrast, the lowest performance is observed when using the lower-precision structures generated by AlphaFold 2.3. These results highlight the importance of structural accuracy in enhancing the predictive power of structure-informed models for antigen-antibody interaction prediction.


\subsection{Integrated Enhancement of HelixFold-Multimer and Energy-Based Method}
Although confidence metrics were not originally developed for interaction predictions, we have demonstrated their relevance to such predictions in prior discussions. In contrast, energy-based methods are specifically engineered for predicting interactions and can deliver accurate results, contingent upon the precision of the input structures. Therefore, investigating the potential for combining these approaches to enhance prediction performance represents a promising and valuable avenue of research.

The integration of ipTM scores from HelixFold-Multimer with FoldX interaction energy scores for binder recognition and affinity prediction is demonstrated in Figures~\ref{fig:fig_inter_main}(e) and (f), respectively.

In Figure~\ref{fig:fig_inter_main}(e), binders (blue dots) and non-binders (red dots) are distinctly separated by their clustering patterns. Binders are predominantly concentrated in the lower-right region, where FoldX interaction energy scores are lower (indicating stronger binding affinities), and HelixFold-Multimer confidence scores are higher. Most binders exhibit negative FoldX interaction energy scores, aligning with the expectation of stronger binding interactions. Additionally, binders are densely distributed in areas with elevated HelixFold-Multimer confidence scores, underscoring the metric’s effectiveness in distinguishing binders from non-binders.
The integration of HelixFold-Multimer confidence scores with FoldX interaction energy scores significantly enhances performance, reducing the false positive rate among the top-ranked candidates compared to using either metric independently. This improvement highlights the complementary roles of the two metrics: FoldX provides detailed energy-based interaction insights, while HelixFold-Multimer enhances structural confidence.

Figure~\ref{fig:fig_inter_main}(f) demonstrates the combined utility of HelixFold-Multimer confidence scores and FoldX energy-based scores in predicting antigen-antibody affinity. High-affinity samples are predominantly located in the region with high structural confidence and low interaction energy (lower right corner), reflecting the expected relationship between these metrics and binding strength.
Notably, the HelixFold-Multimer confidence scores and FoldX energy-based scores exhibit weak correlation across all four antigens, suggesting that each score captures distinct aspects of the binding interaction. HelixFold-Multimer primarily provides structural confidence, while FoldX evaluates energy-based interaction strength. The integration of these complementary metrics offers a more comprehensive and reliable approach to affinity prediction.

\section{Antibody Optimization and Design}

\begin{figure}[h!]
    \centering
    \begin{subfigure}[b]{0.9\textwidth}
        \centering
        \includegraphics[width=\textwidth]{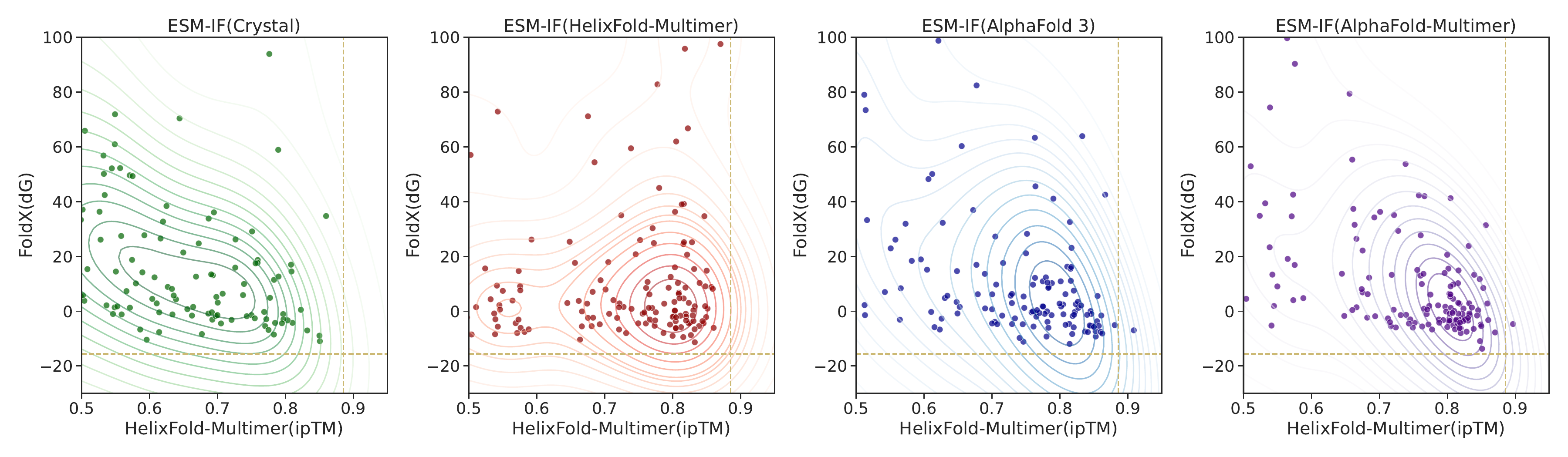}
        \caption{}
        \label{fig:fig_sars_esmif}
    \end{subfigure}
    \begin{subfigure}[b]{0.90\textwidth}
        \centering
        \includegraphics[width=\textwidth]{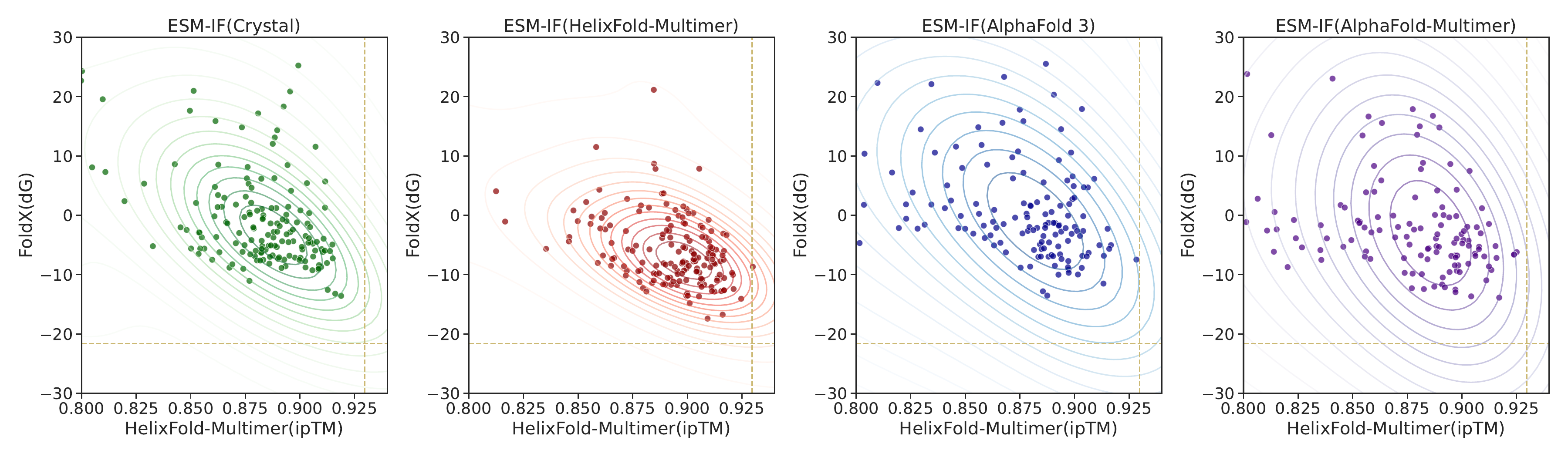}
        \caption{}
        \label{fig:fig_pd1_esmif}
    \end{subfigure}
    \begin{subfigure}[b]{0.9\textwidth}
        \centering
        \includegraphics[width=\textwidth]{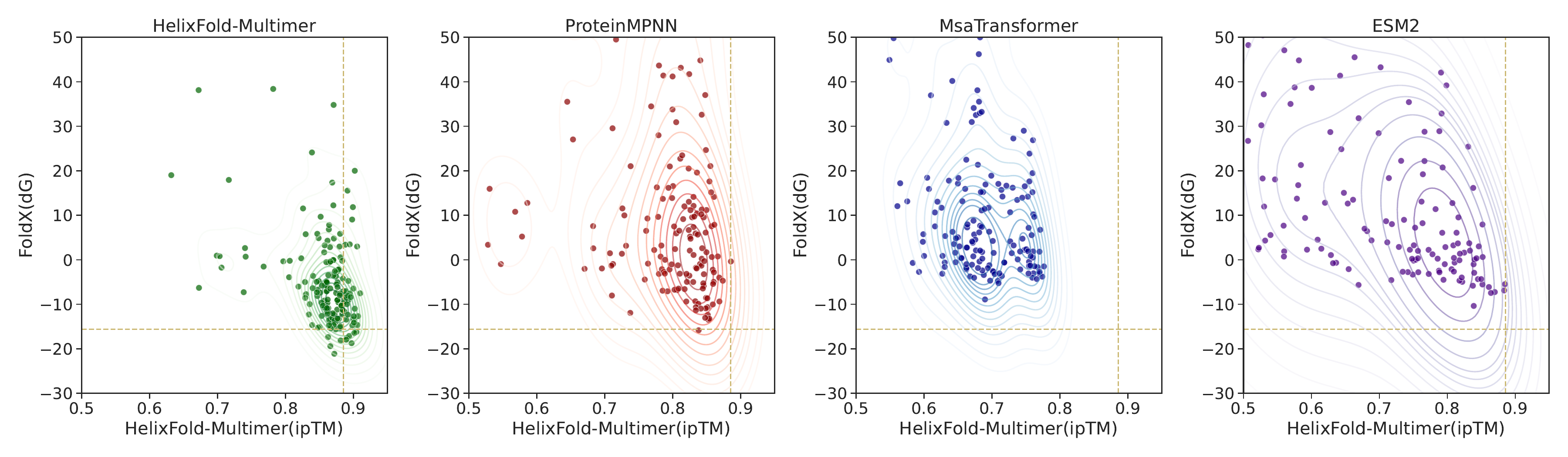}
        \caption{}
        \label{fig:fig_sars_8a96}
    \end{subfigure}
    \begin{subfigure}[b]{0.90\textwidth}
        \centering
        \includegraphics[width=\textwidth]{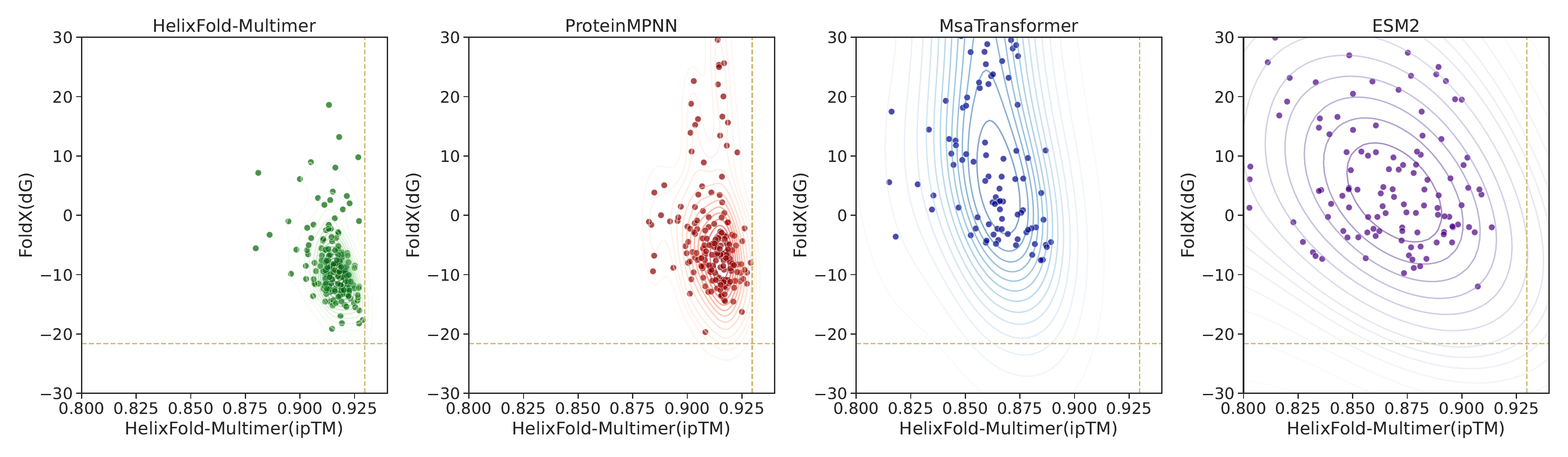}
        \caption{}
        \label{fig:fig_pd1_7wsl}
    \end{subfigure}
    \caption{Comparative Analysis of Antibody Design Performance. (a) and (b) illustrate the performance of ESM-IF with different structural inputs for SARS and PD-1, respectively. (c) and (d) compare the performance of various design methods across targets, SARS and PD-1, respectively. The scatter plots show FoldX scores versus HelixFold-Multimer confidence scores. All the yellow lines mark wild-type antibody positions, serving as benchmarks for design evaluation.}
    \label{fig:design_esmif}
\end{figure}

The design of antibodies with optimized affinities is crucial for therapeutic development. This section explores how HelixFold-Multimer advances antibody design and optimization by leveraging its sophisticated structural prediction capabilities. It also details the use of HelixFold-Multimer's Masked MSA prediction module to generate promising antibody candidates.

\subsection{High-Precision Structures for Better Antibody Design}
In previous sections, we demonstrated that high-precision antigen and reference antibody structures generated by HelixFold-Multimer can be effectively used as inputs for structure-informed models to predict antigen-antibody interactions. Building on this, we are now investigating whether the combination of HelixFold-Multimer with structure-informed models can facilitate the rational design of antibodies. 
Following the methodology outlined in previous studies \cite{shanker2024unsupervised}, we utilized antigen and reference antibody structures predicted by various methods (AlphaFold3, AlphaFold-Multimer, and HelixFold-Multimer) as well as crystallographic structures as inputs for the ESM-IF model. 
This evaluation focused on two common antigens: SARS(Severe Acute Respiratory Syndrome Coronavirus) and PD-1(Programmed Cell Death Protein 1). 
The ESM-IF model was then used to generate several hundred candidate antibodies, which were subsequently evaluated by both ipTM scores and interaction energy scores, as shown in Figure~\ref{fig:design_esmif} (a) (b). 
Antibody sequences generated using HelixFold-Multimer-predicted structures as input for the ESM-IF model exhibit a distribution in scatter plots closer to those obtained with crystallographic structures. This is especially evident for the PD-1 target, where HelixFold-Multimer-predicted designs are closer to the crystallographic results, compared to those from AlphaFold3 and AlphaFold-Multimer. These findings demonstrate the advantage of using more accurate predicted structures to achieve better design outcomes.

\subsection{Masked MSA Prediction Module for Design}
The AlphaFold-style Masked MSA prediction module shares key structural features with BERT-style protein language models, allowing its adaptation to protein sequence design with minimal modifications. This prompted us to investigate whether HelixFold-Multimer's Masked MSA prediction module could effectively redesign antibody CDR regions. To test this, we masked the CDR regions in both the reference antibody sequence and its corresponding MSA, then sample residues for the masked regions based on probabilistic output from the Masked MSA prediction module. This approach generated multiple candidate antibody sequences. Unlike traditional protein language models, HelixFold-Multimer integrates MSA, template structures into its training process, potentially enhancing its ability to perform design tasks. 

We conducted an evaluation of HelixFold-Multimer’s Masked MSA prediction module alongside four baseline methods in two antigens: SARS (PDB: 8a96) and PD-1 (PDB: 7wsl). Notably, the co-crystal structures of these antigens and their corresponding reference antibodies were excluded from the training set of HelixFold-Multimer. For each antigen, 200 candidate antibodies were generated using each method.
Figure~\ref{fig:design_esmif} (c) (d) presents the results of applying HelixFold-Multimer's Masked MSA prediction module to antibody design, compared to four baseline methods on two antigens. Notably, HelixFold-Multimer, MSA Transformer \cite{rao2021msa}, and ESM2 \cite{lin2022language} do not rely on co-crystal structures of antigens and reference antibodies, while ESM-IF \cite{hsu2022learning} and ProteinMPNN \cite{dauparas2022robust} depend on these structures as inputs. The results demonstrate that the Masked MSA prediction module consistently outperforms the baseline methods by generating antibody candidates that predominantly cluster in regions associated with superior properties, such as lower ipTM scores and interaction energy scores. For the SARS antigen, HelixFold-Multimer produced several high-potential candidates that exceeded the reference antibody in both metrics. These findings underscore the potential of the Masked MSA prediction module to advance antibody design methodologies.

\section{Conclusion and Future Work}

Antigen-antibody structure prediction is a critical yet complex task in the development of therapeutic antibodies. Despite the advancements offered by models like AlphaFold3, achieving precise predictions remains a significant challenge. HelixFold-Multimer represents a substantial advancement in this field, markedly improving prediction accuracy through advanced modeling techniques and detailed structural integration. This model not only enhances the precision of structure predictions but also explores new possibilities in antibody development, including interaction prediction and optimization.

Our extensive computational evaluations have demonstrated the effectiveness of HelixFold-Multimer. Looking ahead, we are dedicated to applying HelixFold-Multimer to practical antibody development projects to further refine its capabilities. This ongoing effort aims to maximize its potential and contribute significantly to the advancement of therapeutic antibody research and development.

\clearpage
\section{Method}

\subsection{Datasets}
HelixFold-Multimer was initially trained on an extensive dataset of generic protein complexes sourced from the Protein Data Bank (PDB)\cite{burley2017protein}, comprising entries recorded up to September 30, 2021. In line with the AlphaFold-Multimer approach, the model utilized a self-distillation technique to enhance its predictive capabilities.

Subsequently, structural data of antigen-antibody complexes were curated from the SabDab \cite{dunbar2014sabdab} database. For antibodies, only the fragment variable regions of heavy and light chains were considered. 
HelixFold-Multimer underwent fine-tuning using antigen-antibody data released before January 25, 2023. To prevent overfitting, the training also included data from the general version training dataset and sampling strategies were employed within antigen clusters to ensure a robust and generalized learning process.

The samples from the SAbDab \cite{dunbar2014sabdab} database with release dates between January 25, 2023, and August 9, 2023, are used for evaluation.
We excluded samples with more than 1400 residues. 
The evaluation set comprises 141 antigen-antibody complexes. 
Only the fragment variable regions of the heavy and light chains of the antibodies as inputs for the models.

In addition, antigen residues within a 12 {\AA} distance from the antibody were designated as epitopes and can be incorporated into the model to improve precision. Specifically, if the alpha carbon ($C_{\alpha}$) atom of an antigen residue was within 12 Å of any atom of the antibody, the residue was classified as an epitope residue. HelixFold-Multimer randomly selects several epitope residues as additional input information when epitope information is available.

\label{dataset_antibody}

\subsection{Evaluation Metrics}
\label{sec:Metrics}

\textbf{Structure Prediction:}
To evaluate the accuracy of predicted protein complexes, we employ the DockQ metric \cite{basu2016dockq}. Consistent with common practice, we utilize the DockQ threshold to gauge the success rates of model predictions. Predictions with a DockQ value above 0.23 were deemed accurate, while those surpassing 0.8 were categorized as exhibiting very high accuracy.

\textbf{Interaction Prediction:}
To comprehensively evaluate the binding or non-binding predictions in Antigen-Antibody Interaction, we extend our metrics to include the Area Under the Curve (AUC) of the Receiver Operating Characteristic (ROC) curve, along with Enrichment Factors at 1\% ($\text{EF}^{1\%}$) and 5\% ($\text{EF}^{5\%}$) of the ranked list. These metrics are indicative of the model’s ability to discriminate between binding and non-binding complexes, providing a detailed assessment of its predictive performance across varying degrees of confidence thresholds.

\textbf{Antibody Optimization and Design:}
The HelixFold-Multimer ipTM scores are employed to assess the structural quality of co-crystal complexes between designed antibodies and antigens. Notably, iPTM scores have been validated in interaction prediction studies, demonstrating a correlation with binding affinity. Additionally, the binding affinity of antibody-antigen interactions, calculated using FoldX, serves as a key indicator of the energetic stability and feasibility of the predicted complexes. This metric provides valuable insights into the thermodynamic favorability of the binding process.




\clearpage
\begin{appendices}
\renewcommand{\thesection}{\arabic{section}}
\renewcommand{\thetable}{\arabic{table}}
\renewcommand{\thefigure}{\arabic{figure}}
\setcounter{figure}{0}
\setcounter{table}{0}
\captionsetup[table]{name=Suppl. Table}
\captionsetup[figure]{name=Suppl. Figure}

\section{Additional Results}
\subsection{Structure Prediction}
\begin{figure}[h]
    \centering
    \includegraphics[width=1.0\linewidth]{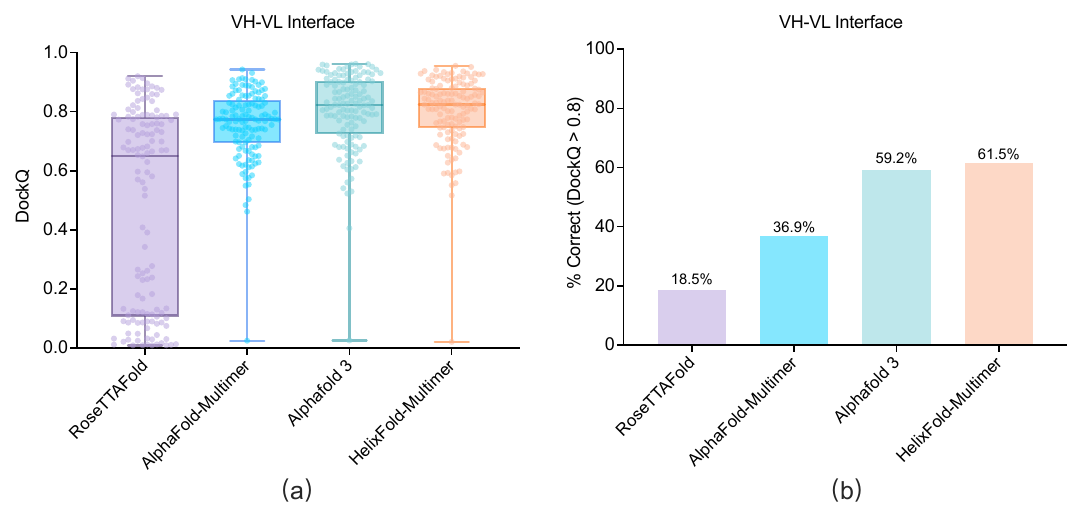} 
    \caption{Prediction accuracy for VH-VL interfaces. (a) DockQ scores for VH-VL antibody interfaces. (b) Percentage of predictions with DockQ scores exceeding 0.8, indicating high-quality interface predictions.}
    \label{fig:fig_ab}
\end{figure}

\begin{figure}[h]
    \centering
    \includegraphics[width=1.0\linewidth]{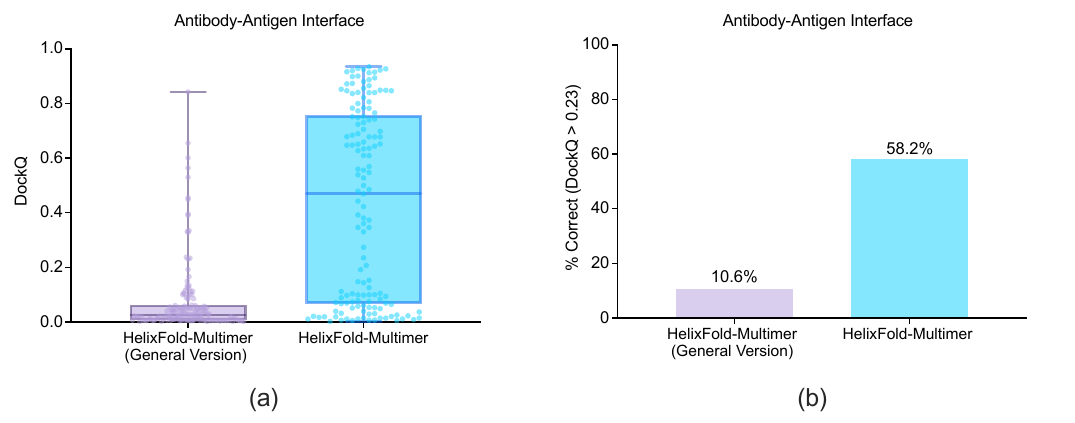} 
    \caption{Performance comparison of the general version and antigen-antibody version of HelixFold-Multimer for antibody-antigen interfaces. (a) DockQ score for antibody-antigen interfaces. (b) Percentage of predictions achieving a DockQ score exceeding 0.23, indicating correct predictions.
    }
    \label{fig:fig_structure_general}
\end{figure}

\begin{figure}[h]
    \centering
    \includegraphics[width=1.0\linewidth]{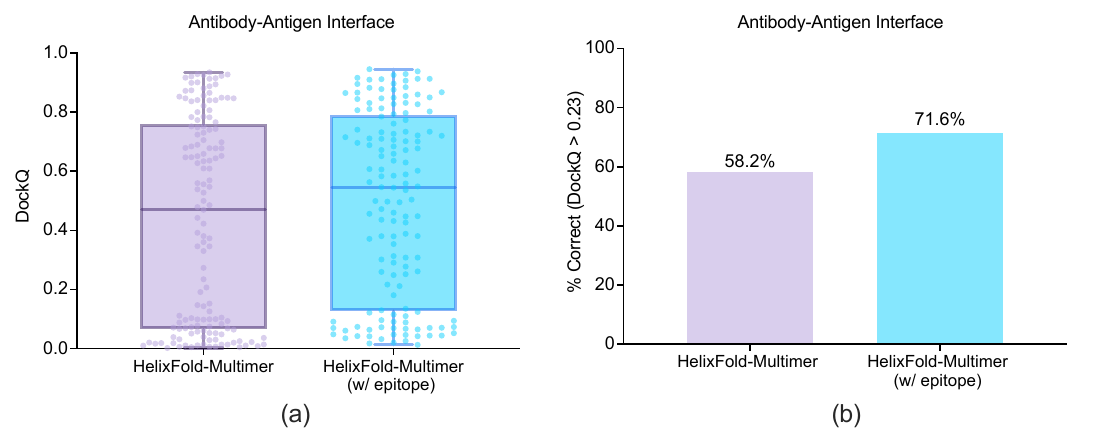} 
    \caption{Impact of epitope specification of HelixFold-Multimer for antibody-antigen interfaces. (a) DockQ scores comparing predictions outputted by HelixFold-Multimer with and without epitope information. (b) Percentage of predictions achieving a DockQ score above 0.23 comparing predictions outputted by HelixFold-Multimer with and without epitope information.}
    \label{fig:fig_epitope_results}
\end{figure}

We additionally evaluated HelixFold-Multimer's structural prediction accuracy for antibody VH-VL interfaces (Suppl. Figure~\ref{fig:fig_ab}), compared the performance of its general version and antibody-antigen specialized version on antibody-antigen interfaces (Suppl. Figure~\ref{fig:fig_structure_general}), and analyzed the effect of incorporating epitope information on prediction accuracy (Suppl. Figure~\ref{fig:fig_epitope_results}).

\subsection{Interaction Prediction}
\begin{figure}[h]
    \centering
    \includegraphics[width=1.0\linewidth]{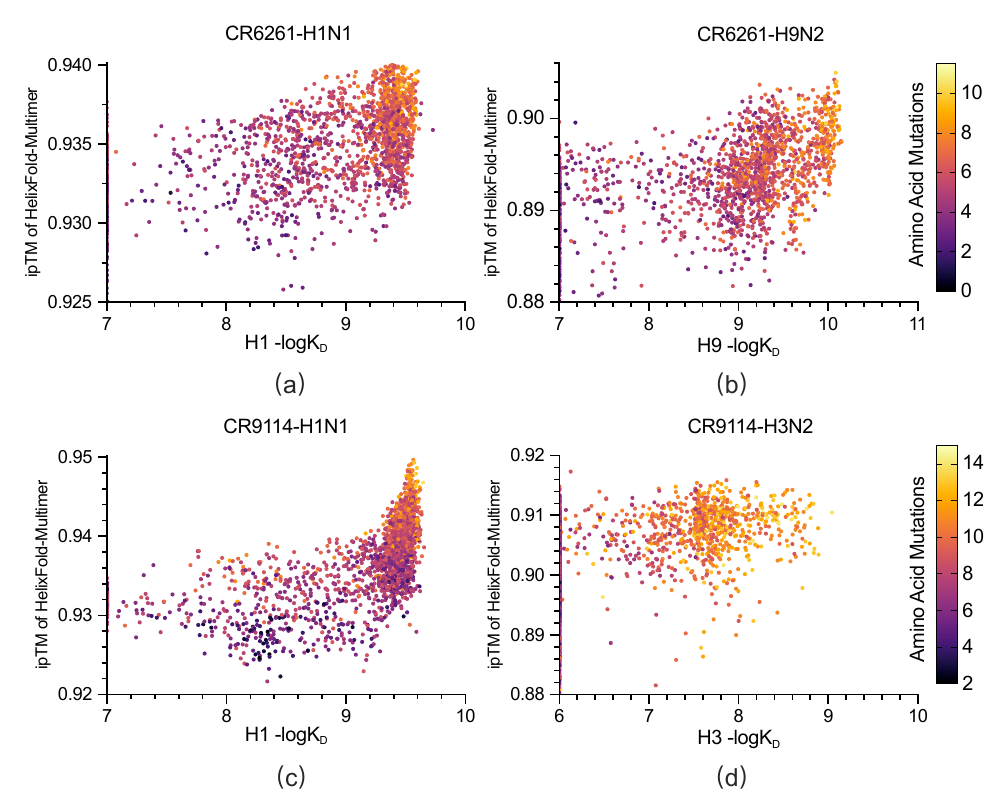} 
    \caption{Correlation between antibody binding affinity and HelixFold-Multimer performance across different influenza antigens. Panels (a) and (b) illustrate the relationship for antibody CR6261 interacting with H1N1 and H9N2 antigens, respectively, while panels (c) and (d) depict the same for antibody CR9114 with H1N1 and H3N2 antigens, correspondingly. The x-axis represents the negative logarithm of the dissociation constant ($-logK_D$), indicating binding affinity, and the y-axis shows the predicted performance metric (ipTM) of HelixFold-Multimer. The color gradient, ranging from purple to yellow, encodes the number of amino acid mutations.}
    \label{fig:fig_dg_affinity}
\end{figure}

We further verified the correlation between HelixFold-Multimer ipTM scores and actual binding affinity, as shown in Suppl. Figure \ref{fig:fig_dg_affinity}, and further evaluated the correlation across different mutation ranges, as presented in Table \ref{tab:Num_mut_3gbn}. The results demonstrated robustness in predicting binding affinity across varying mutation counts, indicating that HelixFold-Multimer's predictions remain consistent and reliable even with significant genetic variations.

We also assessed the effectiveness of integrating structure confidence metrics with energy-based methods in classification tasks across four targets, as shown in Table \ref{tab:bind}, and validated the performance of various computational methods in predicting affinity, as shown in Table \ref{tab:affinity_3gbn}. These findings confirmed that combining HelixFold-Multimer with FoldX enhances prediction accuracy for interactions, highlighting the advantage of integrating structural and energy-based information to effectively capture complex interaction patterns.

The DockQ scores for the predicted structures of PDB IDs 3GBN and 4FQI, generated using HelixFold-Multimer, AlphaFold3, and AlphaFold-Multimer, were assessed. As illustrated in Suppl. Table \ref{tab:dockq_inverse_folding_dg}, these structures were utilized as inputs for the structure-informed language model to facilitate antibody design.

\begin{table}[h]
    \centering
    \setlength{\tabcolsep}{5pt}
    \begin{subtable}[t]{0.46\textwidth}
        \centering
        \begin{tabular}{@{}lSSSS@{}} 
            \toprule 
            {\#Mutations} & {[0,4)} & {[4,6)} & {[6,8)} & {[8,12)} \\ 
            \midrule 
            CR6261-H1N1 & {0.409} & {0.328} & {0.326} & {0.238}   \\
            CR6261-H9N2 & {0.295} & {0.407} & {0.379} & {0.376}  \\
            \bottomrule 
        \end{tabular}
        \caption{Results for CR6261}
        \label{subtab:CR6261}
    \end{subtable}
    \hspace{1.0cm} 
    \begin{subtable}[t]{0.46\textwidth}
        \centering
        \begin{tabular}{@{}lSSSS@{}} 
            \toprule 
            {\#Mutations} & {[0,8)} & {[8,10)} & {[10,12)} & {[12,15)} \\ 
            \midrule 
            CR9114-H1N1 & {0.456} & {0.589} & {0.557} & {0.608}   \\
            CR9114-H3N2 & {0.064} & {0.125} & {0.161} & {0.099}  \\
            \bottomrule 
        \end{tabular}
        \caption{Results for CR9114}
        \label{subtab:CR9114}
    \end{subtable}
    \caption{Correlation between HelixFold-Multimer iPTM scores and binding affinity across various mutation ranges. This table demonstrates the correlation coefficients for the antibodies CR6261 and CR9114 interacting with influenza antigens H1N1 and H9N2 (for CR6261), and H1N1 and H3N2 (for CR9114), across different mutation count ranges. The results confirm the effectiveness of HelixFold-Multimer’s ipTM scores in capturing the correlation with binding affinity consistently across varying degrees of amino acid mutations, highlighting its robust predictive capability in diverse genetic contexts.}
    \label{tab:Num_mut_3gbn}
\end{table}


\begin{table}[h]
    \centering
    \setlength{\tabcolsep}{3pt}
    \begin{tabular}{@{}lSSSSSSSSSSSS@{}} 
        \toprule
        \multirow{3}{*}{Method} & \multicolumn{3}{c}{SARS} & \multicolumn{3}{c}{LYSO} & \multicolumn{3}{c}{VEGF} & \multicolumn{3}{c}{PD-1} \\
        \cmidrule(lr){2-4} \cmidrule(lr){5-7} \cmidrule(lr){8-10} \cmidrule(lr){11-13}
        & AUC & $\text{EF}^{1\%}$ & $\text{EF}^{5\%}$ & AUC & $\text{EF}^{1\%}$ & $\text{EF}^{5\%}$ & AUC & $\text{EF}^{1\%}$ & $\text{EF}^{5\%}$ & AUC & $\text{EF}^{1\%}$ & $\text{EF}^{5\%}$ \\
        \midrule
        {ESM2} & {0.216} & {0.058} & {0.058} & {0.095} & {0.0} & {0.0} & {0.210} & {0.0} & {0.0} & {0.536} & {1.102} & {0.969}\\
        {AlphaFold-Multimer} & {0.581} & {1.579} & {1.464} & {0.407} & {0.0} & {0.0} & {0.533} & {0.0} & {0.132} & \textbf{0.719} & {0.0} & \textbf{5.517} \\
        {HelixFold-Multimer}& \textbf{0.836} & \textbf{9.430} & \textbf{8.173} & \textbf{0.905} & {0.0} & \textbf{0.199} & \textbf{0.959} & {0.0} & \textbf{2.915} & {0.668} & \textbf{19.010} & {5.220}\\
        \bottomrule
    \end{tabular}
    \vspace{5pt}
    \caption{ Comparative analysis of computational methods across multiple protein targets. This table presents the performance metrics for three computational methods: ESM2, AlphaFold-Multimer, and HelixFold-Multimer. Metrics include Area Under the Curve (AUC), Enrichment Factor at 1\% ($\text{EF}^{1\%}$), and Enrichment Factor at 5\% ($\text{EF}^{5\%}$) across four protein targets: SARS, LYSO, VEGF, and PD-1. The results highlight the superior performance of HelixFold-Multimer, achieving the highest AUC and EF values across most targets, demonstrating its enhanced accuracy and reliability in protein-target predictions.}
    \label{tab:bind_compare_method}
\end{table}

\begin{table}[h]
    \centering
    \setlength{\tabcolsep}{3pt}
    \begin{tabular}{@{}lSSSSSSSSSSSS@{}} 
        \toprule
        \multirow{3}{*}{Method} & \multicolumn{3}{c}{SARS} & \multicolumn{3}{c}{LYSO} & \multicolumn{3}{c}{VEGF} & \multicolumn{3}{c}{PD-1} \\
        \cmidrule(lr){2-4} \cmidrule(lr){5-7} \cmidrule(lr){8-10} \cmidrule(lr){11-13}
        & AUC & $\text{EF}^{1\%}$ & $\text{EF}^{5\%}$ & AUC & $\text{EF}^{1\%}$ & $\text{EF}^{5\%}$ & AUC & $\text{EF}^{1\%}$ & $\text{EF}^{5\%}$ & AUC & $\text{EF}^{1\%}$ & $\text{EF}^{5\%}$ \\
        \midrule
        {FoldX} & {0.807} & \textbf{11.520} & {8.382} & {0.959} & \textbf{4.138} & {4.392} & {0.957} & {4.721} & {6.360} & {0.578} & {11.410} & {6.712}\\
        {HelixFold-Multimer}& {0.836} & {9.430} & {8.173} & {0.905} & {0.0} & {0.199} & {0.959} & {0.0} & {2.915} & \textbf{0.668} & {19.010} & {5.220}\\
        {HelixFold-Multimer + FoldX}& \textbf{0.846} & {10.470} & \textbf{8.802} & \textbf{0.972} & {3.621} & \textbf{4.591} & \textbf{0.985} & \textbf{6.745} & \textbf{6.757} & {0.663} & \textbf{19.010} & \textbf{6.712}\\
        \bottomrule
    \end{tabular}
    \vspace{5pt}
    \caption{Comparative analysis of computational methods across multiple protein targets. This table presents the performance metrics for three different computational methods: FoldX using HelixFold Structure, HelixFold-Multimer alone, and a combined approach of HelixFold-Multimer with FoldX. The metrics shown are Area Under the Curve (AUC), Enrichment Factor at 1\% ($\text{EF}^{1\%}$), and Enrichment Factor at 5\% ($\text{EF}^{5\%}$) across four different targets: SARS, LYSO, VEGF, and PD-1. The results underscore the effectiveness of HelixFold-Multimer, and further indicate that combining it with an energy-based method like FoldX can enhance prediction accuracy.}
    \label{tab:bind}
\end{table}

\begin{table}[h]
    \centering
    \setlength{\tabcolsep}{5pt}
    \begin{tabular}{@{}lSSSS@{}} 
    \toprule 
    {Method} & {CR6261-H1N1} & {CR6261-H9N2} & {CR9114-H1N1} & {CR9114-H3N2} \\ 
    \midrule 
    FoldX  & {0.482} & {0.379} & {0.536} & {\textbf{0.313}}  \\
    HelixFold-Multimer & {0.540} & {0.528} & {0.575} & {0.188}   \\
    HelixFold-Multimer + FoldX & {\textbf{0.617}} & {\textbf{0.596}} & {\textbf{0.619}} & {0.241}   \\
    \bottomrule 
    \end{tabular}
    \hspace{1cm}
    \vspace{5pt}
    \caption{Comparison of Pearson correlation coefficients (PCC) for Different computational methods across antibody-antigen interactions. This table displays the PCC metrics, which quantify the correlation between observed and predicted data, for three computational methods: FoldX (using HelixFold structure), HelixFold-Multimer, and a combination of HelixFold-Multimer with FoldX. Results are shown for interactions between antibodies CR6261 and CR9114 with influenza antigens H1N1 and H9N2 (for CR6261) and H1N1 and H3N2 (for CR9114).}
    \label{tab:affinity_3gbn}
\end{table}

\begin{table}[h]
    \centering
    \setlength{\tabcolsep}{5pt}
    \begin{tabular}{@{}lSSS@{}} 
    \toprule 
    {Method} & {HelixFold-Multimer} & {AlphaFold 3} & {AlphaFold-Multimer} \\ 
    \midrule 
    {3GBN} & {0.910} & {0.555} & {0.188}  \\
    {4FQI} & {0.848} & {0.609} & {0.086}  \\
    \bottomrule 
    \end{tabular}
    \hspace{1cm}
    \vspace{5pt}
    \caption{Comparative DockQ scores for predicted structures of PDB IDs 3GBN and 4FQI using HelixFold-Multimer, AlphaFold 3, and AlphaFold-Multimer. The scores highlight the differing accuracies of these methods, with HelixFold-Multimer consistently outperforming the others in this evaluation. }
    \label{tab:dockq_inverse_folding_dg}
\end{table}

\clearpage
\section{Detailed Methods}
\subsection{Data Pipeline}

The data pipeline is the first step when running HelixFold-Multimer. It takes an input an mmCIF/PDB file and produces input features for the model.

\subsubsection{Parsing}
Following the methodology of AlphaFold-Multimer, we parse mmCIF data, extracting fundamental metadata including resolution, release date, and methodological details, alongside non-coordinate information such as bioassembly specifics, chemical component particulars, chain names and sequences, as well as covalent bond data. For the optional input of epitope information of antigen-antibody complexs, we also parse it from the structural data, selecting antigenic residues within a 12 {$\AA$} distance of the alpha carbon atoms of antibodies as epitopes. Additionally, for antigen-antibody data, we employ the ANARCI\cite{dunbar2016anarci} tool to determine the presence of antibody chains.

\subsubsection{MSA Search}
We conducted MSA searches using the following search tools and datasets (see table \ref{tab:tools}):

{Jackhmmer searches use the following additional flags:}
-N 1 -E 0.0001 \text{--incE} 0.0001 \text{--F1} 0.0005 \text{--F2} 0.00005 \text{--F3} 0.0000005.

{HHBlits searches use:} -n 3 -e 0.001 \text{--realign\_max} 100000  \text{--maxfilt} 100000 \text{--min\_prefilter\_hits} 1000 -p 20  -Z 500.

The following databases were used:
For training and inference models, we searched UniRef90\cite{suzek2015uniref} (up to v2022\_05), UniProt\cite{uniprot2023uniprot} (v2020\_05), Reduced BFD\cite{tunyasuvunakool2021highly}, Uniclust30\cite{mirdita2017uniclust} (v2018\_08) and MGnify\cite{mitchell2020mgnify} (up to v2022\_05).

\begin{table}[h]
    \centering
    \setlength{\tabcolsep}{5pt}
    \begin{tabular}{@{}llll@{}} 
    \toprule 
    {Database} & {Search tool} & {Database-specific flags} & {Max sequences}  \\ 
    \midrule 
    UniRef90 & jackhmmer & --seq\_limit 100000 & 10000    \\
    UniProt & jackhmmer & --seq\_limit 500000 & 50000   \\
    Uniclust30 & HHBlits v3.0 &  & None\\
    Reduced BFD & jackhmmer & --seq\_limit 50000 & 5000    \\
    MGnify & jackhmmer & --seq\_limit 50000 & 5000   \\
    \bottomrule 
    \end{tabular}
    \caption{Search tool configuration for various protein sequence databases. This table lists the configurations used with different protein sequence databases, detailing the search tool, any database-specific flags, and the maximum number of sequences allowed in a search. The data covers five major databases: UniRef90, UniProt, Uniclust30, Reduced BFD, and MGnify, with specific flags tailored to optimize search queries in jackhmmer for most databases and HHBlits v3.0 for Uniclust30.}
    \label{tab:tools}
\end{table}

The UniProt search result was kept separate and used to provide cross-chain genetic information, as in [18]. All the results except UniProt were stacked in the order of UniRef90, Reduced BFD, MGnify and deduplicated to form the main MSA.
\subsubsection{Training Data}
\textbf{PDB}: The targets released before 2020-05-14 in PDB are used to train HelixFold-Multimer. We filtered out the targets with a resolution higher than 3Å and those containing less than 10 amino acids. The targets are clustered at 40\% sequence identity cutoff by mmseq2.

\textbf{Distillation-Uniclust30}: We inference the structures of the targets in Uniclust30 (version 2018-08) by AlphaFold-Multimer. We follow the data-prepossess procedure reported in AlphaFold-Multimer. Further, the target structures with an average pLDDT of less than 50 are filtered out. Then, the targets are clustered at 30\% sequence identity cutoff.

\textbf{Distillation\-EBI}: About one million protein structures are extracted from AlphaFold Protein Structure Database v2 (version 2022\-01)\cite{varadi2022alphafold}. We removed the protein structures with an average pLDDT of less than 50. The remaining targets are clustered at 50\% sequence identity cutoff.

\textbf{SabDab}: The SabDab structures released before  January 25, 2023 were used to train HelixFold-Multimer. we employ the mmseqs\cite{steinegger2017mmseqs2} tool to cluster the longest antigen chains with an identity cutoff of 80\%. we excluded samples containing more than 1400 residues. We extracted only the fragment variable regions of the heavy and light chains of the antibodies as inputs for the model.

\subsubsection{Evaluation Data}
\label{evaluation_data}
\textbf{Structure Prediction}:
We selected samples from the SAbDab \cite{dunbar2014sabdab} database with release dates between January 25, 2023, and August 9, 2023, as the evaluation set for antibody-related data, ensuring that the evaluation samples are not present in our training set. Further, the antigen type of peptide and protein are chosen. Additionally, we exclude samples containing more than 1400 residues. 
In cases where AlphaFold 3 predictions failed, we intersect the successful predictions from both AlphaFold 3 and HelixFold-Multimer.
The test set comprises 141 antigen-antibody complexes. The antibody chains from antigen-antibody complexes are extracted to serve as a test dataset for the evaluation of antibody variable regions (VH-VL). We extract only the fragment variable regions of the heavy and light chains of the antibodies as inputs for the models.

The antigen-antibody test PDBs: 7u9e 8ium 8iw9 8dzv 8dy1 8iuk 8dy5 8axh 7yue 7ua2 8cz8 8db4  8hc4  8epa  8hc5  8hhy  8hhx  7uxl  7xdk  7xcz  7xda  8ek1  8eka  7xdl  8c3v  8bcz  8d7e  7wnb  7wn2  8i5i  8saw  7umn  7xj6  8dto  8sb0  8sb1  8saq  8sar  7zjl  8say  8i5h  8sb5  8sas  8sb3  8sb2  7uja  7uow  8ct6  8sav  7xj8  7xj9  8sax  7xik  8heb  7xs8  8hec  7xil  8hed  8a96  8gs9  8f0h  8fax  8fg0  8a99  7xsa  7xrz  8bse  7yru  8bsf  7zqt  8dn6  8hn6  8hn7  7st5  8dn7  8f6o  8gb8  8gb6  8ahn  8f60  8cwi  8cwj  8gb7  8dnn  8cwk  7xsc  8f6l  8av9  8g3n  8g3v  8g3r  7xeg  7xjf  8gnk  8g3q  8ee1  8ee0  8g3p  8g3o  8dwy  8g3z  8g3m  8dww  7quh  8g30  8h07  8czz  7xei  8a44  8d9y  8eoo  8da0  8da1  8d9z  8e1m  8scx  8smt  7zoz  8e1g  8duz  8elo  8elq  8de4  8cim  8elp  8gtp  7yd1  8byu  8gtq  7y8j  8el2  8ol9  7yds  8ib1  7yv1  8dz3  8dyx  7trh  8e6j  7yk4  8e6k

\textbf{Interaction Prediction}:
Firstly, we collected binding data for several targets: SARS-CoV2 receptor-binding domain (RBD), Lysozyme, vascular endothelial growth factor(VEGF) and programmed cell death protein 1 (PD-1). To assess the model's capability in differentiating between binding and non-binding antibodies, we created a non-binding antibody dataset comprising 1,000 random antibodies from a healthy donor.  This dataset was applied accoss four targets. 
The SARS-CoV-2 dataset was derived from a single B cell sequencing dataset obtained from the AS database \cite{olsen2022observed}, yielding a set of 128 antibodies by random select.
The PD-1 dataset included 27 therapeutic antibodies sourced from Thera-SAbDab \cite{raybould2020thera}.
The anti-Lysozyme dataset was designed to evaluate the mutational tolerance within the original antibody D44.1, 200 mutated antibodies were selected randomly\cite{chungyoun2023flab}.
Lastly, the VEGF dataset consisted of 150 mutational variants, each spanning the highly variable region of the anti-VEGF antibody (G6.31)\cite{chungyoun2023flab}.

Secondly, we gathered datasets of antibody affinity for influenza strains (H1N1, H3N2, H9N2) from the literature \cite{phillips2021binding}. This dataset includes data for two mutated antibodies: CR6261 and CR9114. For CR6261, we used all available affinity data against H1N1 and H9N2, with approximately 2000 samples for each target. For CR9114, to reduce computational complexity, we sampled 2000 samples for each target based on the distribution of the affinity data. For the H1N1 affinity data, we randomly sampled 2000 samples that replicate the overall distribution of affinity across all data. For H3N2, due to its highly imbalanced affinity data distribution, where high-affinity samples are less frequent, we applied a 5:1 weighting to more frequently sample high-affinity examples, ensuring some degree of sample balance. Ultimately, we obtained about 2000 samples for each type of affinity data for CR6261 (H1N1, H9N2) and CR9114 (H1N1, H3N2).

\textbf{Antibody Design}:
we selected three widely studied antigen targets: SARS, PD-1. The specific reference antibody complexes used were SARS (PDB ID: 8a96), PD-1 (PDB ID: 7wsl), none of which were present in the training set.

\end{appendices}

\clearpage

\bibliographystyle{unsrt}  
\bibliography{references}

\end{document}